# Prospective Action-Resource Graph Multi-Agent Reinforcement Learning for Electric HGV Fleets under Shared Charging Infrastructure

**Yiming Xu[1, *], Fulong Yao[2], Maurizio Albano[1], Rigel Gjomemo[3], Liana Cipcigan[1]**

*1. School of Engineering, Cardiff University, CF24 3AA, United Kingdom*

*2. School of Computer Science and Informatics, Cardiff University, CF24 4AG, United Kingdom*

*3. College of Engineering, University of Illinois at Chicago, IL 60607, United States*

## Abstract

Electric heavy goods vehicle (eHGV) fleet coordination requires each vehicle to choose driving, charging and resting actions whose consequences depend on shared infrastructure. Existing multi-agent reinforcement learning (MARL) methods mainly represent current states, leaving action-specific futures and resource interactions to be learned indirectly. We develop Horizon-Preview and Graph-Coordinated Proximal Policy Optimisation (HGC-PPO), which integrates horizon preview and candidate-resource graph coordination into the PPO decision architecture. Horizon preview represents each legal action through its short-horizon operational consequences. Graph message passing then updates these candidates through station, charger-type and time-slot resources shared across the fleet. On the 794km Folkestone-Glasgow task, HGC-PPO achieved the lowest mean charging cost among five RL methods at £504.32 per vehicle, with a fleet completion time of 24.920h. It also completed the task 1.763h faster and cost £36.51 less per vehicle than a replay-verified finite-budget MILP benchmark. Across six unseen routes and fleet sizes from 5 to 20 vehicles, HGC-PPO delivered the lowest macro-average completion time, charging cost and charging frequency among the RL methods. A $2\times2$ ablation showed that the full preview-graph architecture performed best across all four operational metrics. These results show that prospective action information and fleet-wide resource coordination improve eHGV decisions under charging and driver-hours constraints.

---

*Corresponding author: Yiming Xu
Email address: XuY130@cardiff.ac.uk (Yiming Xu[1]), YaoF@cardiff.ac.uk (Fulong Yao[2]), AlbanoM@cardiff.ac.uk (Maurizio Albano[1]), rgjome1@uillinois.edu (Rigel Gjomemo[3]), CipciganLM@cardiff.ac.uk (Liana Cipcigan[1])

# 1. Introduction

Road-freight decarbonisation increasingly depends on electric heavy goods vehicles (eHGVs), because HGVs produce up to 20% of UK domestic transport greenhouse gas emissions despite representing a small share of road vehicles [1], [2]. In response, phase-out targets for non-zero-emission HGVs have been introduced in the United Kingdom, alongside public programmes supporting zero-emission trucks and charging infrastructure [3], [4], [5]. These measures support fleet electrification, but they do not remove the operational difficulty of long-haul deployment. Route progression, battery energy, charging opportunities and statutory driving limits must still be managed jointly. Under UK drivers' hours rules [6] , mandatory breaks are required after continuous driving, together with longer daily rest periods, while charging can sometimes overlap with these rests. When several eHGVs travel along the same corridor, their charging decisions become interdependent: selecting a station or charging time changes charger occupancy, queueing and the options available to other vehicles [7], [8]. Charging is therefore a fleet-level shared-resource decision rather than an independent energy-management task [9], [10].

Rule-based, optimisation-based and reinforcement-learning methods provide complementary approaches to this problem. Rule-based policies offer transparent operational logic but do not learn interaction-dependent decisions from changing fleet states [11]. Mixed-integer linear programming can produce rigorous schedules and optimisation bounds for fixed cases [12], [13], although a formulation that simultaneously represents continuous-time queueing, shared charger capacity, vehicle energy, charging-rest overlap and regulation-dependent decisions can become computationally demanding and must be resolved as operating conditions change. Reinforcement learning (RL) provides fast policy execution after training [14], [15], while multi-agent reinforcement learning (MARL) allows vehicles to learn within a shared environment [7], [16], [17]. Independent policies, however, do not directly represent how one vehicle's prospective action changes the resource conditions faced by others. Centralised critics and shared rewards improve coordination during learning [18], [19], but they do not by themselves make the resource consequence of every candidate action explicit in the policy representation.

Recent graph-based MARL and EV charging-guidance methods have demonstrated the value of relational representations for modelling vehicles, transportation networks, charging stations and anticipated charging demand [8], [20], [21] . A remaining methodological question is how to represent every currently feasible action, rather than only agents or physical entities, as a prospective decision object, and how to revise that action representation according to other vehicles' potential use of the same resource before any action is executed. This question is particularly important for eHGV fleets because a charging or rest decision affects several coupled outcomes, including fleet completion time, charging expenditure, queueing and subsequent charging requirements. A policy that observes only the current system state must infer these action-level consequences indirectly through repeated interaction and delayed returns. By making prospective consequences explicit, a direct coordination signal can be provided before action selection instead of relying exclusively on delayed returns.

To address this challenge, we develop Horizon-Preview and Graph-Coordinated Proximal Policy Optimisation (HGC-PPO), an integrated extension of PPO that introduces prospective coordination directly into the MARL decision architecture. At each decision epoch, horizon preview evaluates every legal action in an isolated operational state and represents its short-horizon consequences as a candidate-action node. These nodes are connected to the station, charger-type and time-slot resources that they may use, allowing bidirectional graph message passing to propagate prospective fleet demand and return the resulting resource information to each candidate before action selection. A parameter-shared actor evaluates the coordinated candidate representations, while a fleet-level critic supports joint PPO training. Together, these operations extend the policy's decision context from the current system state to action-specific operational futures, the next compulsory service boundary and the prospective resource demand created by other vehicles.

This study makes three contributions. First, HGC-PPO expands the information available to MARL policy evaluation. The policy can distinguish not only which actions are currently legal, but also how far each action advances the vehicle, how it changes SOC, time, charging, queueing and rest requirements, which compulsory behaviour follows, and how the resulting resource demand interacts with the prospective choices of other vehicles. This information supports fleet decisions that account for future operational consequences and shared-infrastructure competition before execution. Second, the study formulates eHGV fleet operation as a regulation- and infrastructure-coupled multi-agent decision problem in which driving, charging, queueing, mandatory rest, charging-rest overlap, finite charger capacity and dynamically changing legal actions evolve on a unified operational timeline. Third, the proposed RL method is evaluated through matched comparisons with PPO, IPPO, MAPPO and HAPPO, together with a rule-based policy and replay-verified optimisation benchmarks. The evaluation combines three independent training seeds, six unseen routes, four fleet sizes, a complete 2×2 architectural ablation and vehicle-resource behaviour analysis.

The remainder of this paper is organised as follows. Section 2 reviews optimisation, reinforcement-learning and graph-based approaches to electric-vehicle operation and charging coordination. Section 3 presents the regulation- and infrastructure-coupled eHGV fleet model. Section 4 introduces HGC-PPO, including horizon preview, candidate-resource graph coordination, policy evaluation and PPO learning. Section 5 describes the data, benchmarks and experimental protocol. Section 6 presents the training comparison, cross-route and fleet-load generalisation, ablation and behavioural analyses. Section 7 discusses the combined implications and operating boundaries of the results. Section 8 concludes the paper.

# 2. Related work

Electric-vehicle routing was initially formulated by extending classical vehicle-routing models with battery limits, charging stops and time-window constraints [11], [12], [13], [22]. Greater operational detail was subsequently introduced through partial recharging [23], nonlinear charging functions [13], shared charging stations [24], capacitated charging infrastructure [25], [26] and anticipatory routing under public-station congestion [27]. At the network-planning level, charging-station locations in congested intercity networks were optimised by Bao and Xie [28] , whereas the present study addresses real-time fleet coordination under a given charging infrastructure. These models established rigorous representations of route feasibility and charging operations. However, vehicle interactions are generally resolved through a centrally constructed schedule or predefined routing policy. When charger availability, queueing and vehicle states evolve during operation, the optimisation problem must be repeatedly resolved or approximated through additional assumptions.

Truck-specific constraints have also been incorporated into electric-freight planning. Joint routing and charging schedules were formulated for eHGV delivery operations in [29], while long-haul charging flexibility and alternative charging opportunities were considered in [30]. Charging was coordinated with statutory rest requirements through the break-and-charge formulation in [31] and routing, charging, platooning and driver assignment were jointly represented in [32]. Bai et al. addressed the closest problem settings to the present study. Their 2023 work formulated hours-of-service-aware charging for a single electric truck travelling on a pre-planned route and developed a rollout approximation for real-time decision-making [9]. Their subsequent study coordinated multiple electric trucks under limited charging facilities and travel uncertainty through station-provided waiting-time estimates and distributed charging-plan updates [10]. These studies establish that driver-hours constraints and capacity-limited charging can be incorporated into online electric-truck decisions. Their methods optimise or update truck-level charging plans, whereas HGC-PPO learns a fleet policy that constructs every legal action as a prospective candidate and coordinates these candidates through shared station–charger–time resources before action selection. The distinction lies in embedding action-conditioned operational futures and prospective fleet-wide resource competition directly into PPO policy evaluation.

RL has been applied where charging and routing decisions must be made sequentially under uncertain system conditions. Fleet charging control, dynamic EV routing and the broader design space of RL-based charging scheduling have been studied in [33], [34], [35]. Multi-agent formulations have subsequently been used to coordinate charging demand, vehicle schedules and transport–energy interactions [7], [16], [17], [36]. Shared charging and refuelling capacity have also been considered for port drayage trucks through COMA-based MARL [37]. In these approaches, coordination is commonly induced through shared rewards, centralised critics, attention mechanisms or information exchanged among agents. These mechanisms allow interactions to be learned, but the operational consequence of each currently available vehicle action is generally not represented as an explicit object before that action is selected.

PPO provides a widely used policy-optimisation mechanism based on clipped updates [38]. Its cooperative multi-agent variants include IPPO and MAPPO [18], while sequential agent-wise updates are employed in HAPPO [19]. These methods differ in how policy parameters, critics and agent updates are organised, but none prescribes how charging-resource consequences should be represented in the policy input. They are therefore treated in this study as matched policy-learning baselines. HGC-PPO retains the PPO update mechanism; its methodological contribution lies in the representation and coordination performed before the actor evaluates the available actions.

Graph-based reinforcement learning has provided a more explicit representation of relationships among vehicles, charging stations, road networks and electricity-system entities. Heterogeneous road, charging-station and power-network relations were encoded through graph attention in [8]. A scalable EV–charger–transformer representation was developed to support large charging systems in [20], while anticipated charging demand and interactions among charging-station agents were incorporated into charging-guidance models in [21] and [39] . These studies demonstrate that graph structure can improve relational reasoning, scalability and charging coordination. Their graphs are primarily formed from physical entities, network topology, present system states or forecast charging demand. By contrast, an unexecuted action is not generally constructed as a node whose short-horizon operational consequence is revised through the resources that may also be selected by other vehicles.

The remaining gap is therefore not the absence of MARL or GNNs in electric-vehicle operation. It is the absence of an action-level prospective representation that jointly answers three questions before execution: which actions are currently legal, what operational consequence is expected from each action, and how that consequence may change when other vehicles depend on the same resource. HGC-PPO addresses this gap by constructing a prospective candidate–resource graph. Each legal action is represented as a candidate-action node with a short-horizon consequence vector; candidate-to-resource and resource-to-candidate message passing then produces coordinated candidate representations for policy evaluation. This design makes prospective shared-resource interactions available to the actor directly, rather than requiring them to be inferred only from current-state embeddings, shared rewards or delayed returns.

# 3. Framework and system modelling

## 3.1. Framework overview

Figure 1 presents the overall framework for coordinated eHGV fleet operation. Four coupled layers are included: the real-world data input layer, route generation layer, behaviour and regulation coupling layer, and HGC-PPO decision layer. Together, these layers convert transport, infrastructure and regulatory data into feasible vehicle actions and coordinated fleet decisions.

The real-world data input layer integrates the UK road network, charging-station locations and capacities, electricity prices, eHGV specifications, UK drivers' hours rules and scenario-specific infrastructure restrictions. Road and charging data are processed by the route generation layer, where HGV-compatible road segments are retained and unavailable roads or charging facilities are excluded for the specified operating scenario. Feasible route nodes and charging opportunities are then supplied to the operational environment.

The behaviour and regulation coupling layer governs the physical execution of driving, charging, queueing and resting activities. Battery feasibility, charger capacity and the UK assimilated drivers' hours rules [6] are enforced when the legal action set is constructed. Charging and rest can share the same non-driving period when the operational conditions defined in Section 3.3 are satisfied.

At each decision epoch, the environment supplies vehicle states and legal action sets to the HGC-PPO decision layer. Horizon preview represents the short-horizon consequence of each legal action as a candidate-action node, after which graph message passing coordinates candidates that may depend on the same resource. The coordinated candidate representations are evaluated by a shared actor to produce vehicle actions. Their execution updates vehicle, charger and regulatory states, thereby closing the interaction loop. During training, fleet-level information is evaluated by the critic and the differentiable policy components are updated through PPO, as detailed in Section 4.

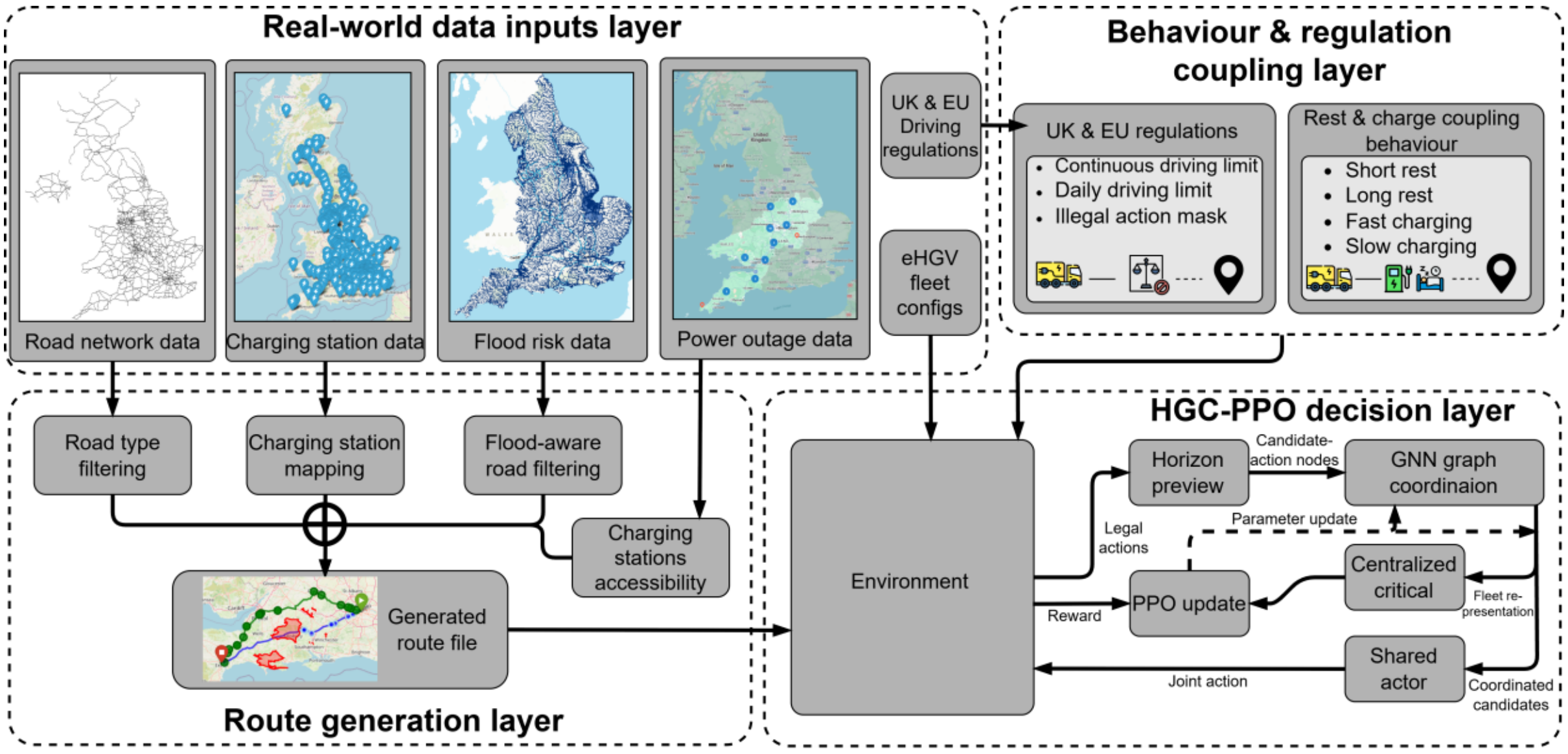

Figure 1. Overall framework for coordinated eHGV operation with the HGC-PPO decision layer.

## 3.2. Event-driven multi-agent problem formulation

Coordinated eHGV operation is formulated as an event-driven multi-agent Markov decision process. Each vehicle is treated as an agent, while charging stations constitute shared resources whose availability is jointly affected by fleet decisions. Decision epochs occur at route nodes where an active vehicle must select a driving, charging or resting action. The environment first determines the legal actions according to vehicle energy, charger accessibility and driver-hours conditions; the policy then selects among these actions. The problem is evaluated through fleet completion time, charging expenditure, queueing time and charging frequency, subject to the feasibility rules defined in Section 3.3.

3.2.1. State definition

Let the fleet contain $N$ eHGVs. Their index set $\mathcal{I}$ is:

$$I = \{1, 2, \dots, N\} \tag{1}$$

The operational state of vehicle $i$ in $\mathcal{I}$ at event-based decision step $t$ is:

$$s_i^t = (p_i^t, soc_i^t, \tau_i^t, h_{i,c}^t, h_{i,d}^t, r_{i,s}^t, r_{i,l}^t, \varepsilon_i^t) \tag{2}$$

Here, $p_i^t$ denotes the current route or charging-station node, $soc_i^t$ is the battery state of charge, and $\tau_i^t$ is elapsed operational time. The variables $h_{i,c}^t$ and $h_{i,d}^t$ record continuous and daily driving time. The variables $r_{i,s}^t$ and $r_{i,l}^t$ record uninterrupted non-driving time towards the short- and long-rest requirements. The completion indicator $\varepsilon_i^t$ equals one after the destination is reached and zero otherwise. The superscript $t$ indexes decision events, which need not occur at the same elapsed time for every vehicle.

The corresponding fleet-level state $s^t$ is:

$$s^t = (s_1^t, s_2^t, \dots, s_N^t, q^t, \xi^t) \tag{3}$$

$q^t$ is the shared charging-resource state, including charger occupancy and the next availability time of each fast or slow charger. The term $\xi^t$ contains route and infrastructure information, including inter-station distances, charger capacities, charging powers, electricity prices, and scenario-specific facility availability.

3.2.2. Observation definition

Each vehicle receives an 18-dimensional raw observation $o_i^t = \varphi_i(s^t)$, where $\varphi_i$ is the deterministic observation map for vehicle $i$. The observation contains six feature groups: (1) current SOC and remaining route distance; (2) continuous and daily driving time and their remaining legal allowances; (3) sine and cosine encodings of time of day; (4) current-station waiting times and charging demand-to-capacity ratio; (5) downstream SOC, driving-time, queue, and charging-demand information; and (6) accumulated blocked time and the normalised fleet-group identifier. All continuous variables are normalised before policy evaluation. The complete feature definitions are provided in Section S1 of the Supplementary Material.

This raw observation should be distinguished from the candidate-action representation introduced in Section 4. The observation describes the current vehicle and infrastructure state. Horizon preview subsequently combines that state with each legal action to construct its consequence vector and candidate-action node.

3.2.3. Action definition

The joint fleet action $a^t$ at decision step $t$ is:

$$a^t = (a_1^t, a_2^t, \dots a_N^t), a_i^t \in \mathcal{A}_i^t \tag{4}$$

The common discrete action space is:

$$\mathcal{A} = \{\text{drive}_k, \text{charge}_k | k = 1, \dots, K\} \cup \{\text{rest}\}, K = 3. \tag{5}$$

Here, $\text{drive}_k$ advances the vehicle to downstream candidate station $k$ without charging; $\text{charge}_k$ advances it to the same candidate and starts an available charging service; and rest keeps the vehicle at its current node for the required non-driving period. With $K = 3$, the common action space contains seven actions.

The set $\mathcal{A}$ defines the common action vocabulary, whereas $\mathcal{A}_i^t$ is the legal set for vehicle $i$ at decision step $t$. Energy-, infrastructure-, or driver-hours-infeasible actions are removed before policy evaluation. HGC-PPO therefore ranks only actions that the environment can execute.

3.2.4. Operational objectives and feasibility

For policy $\pi$, operational performance is represented by the four-component outcome vector $m(\pi)$:

$$\text{m}(\pi) = \text{E}_\pi[T_{\text{fleet}}, \bar{C}, \bar{Q}, \overline{N_{\text{chg}}}]. \tag{6}$$

The components of the outcome vector are defined as:

$$\begin{gathered} T_{\text{fleet}} = \max_i \tau_i^{\text{arr}}, \bar{C} = \frac{1}{N} \sum_{i \in I} C_i, \\ \bar{Q} = \frac{1}{N} \sum_{i \in I} Q_i, \overline{N_{\text{chg}}} = \frac{1}{N} \sum_{i \in I} n_i^{\text{chg}}. \end{gathered} \tag{7}$$

$T_{\text{fleet}}$ is fleet completion time, defined by the latest vehicle arrival time $\tau_i^{arr}$. The quantities $C_i$, $Q_i$, and $n_i^{chg}$ are the total charging expenditure, cumulative charger-queueing time, and number of charging events for vehicle $i$. Their fleet means are denoted by $C$, $Q$, and $N_{\text{chg}}$, respectively. $\text{E}_\pi$ denotes expectation under policy $\pi$. The four outcomes are reported separately because no externally validated utility function makes them interchangeable.

Policy execution remains subject to battery feasibility, charger accessibility, finite charger capacity and driver-hours requirements. Their state transitions and feasibility conditions are specified in Section 3.3. The PPO training return and the fixed checkpoint-selection rule are introduced separately in Sections 4 and 5; neither replaces the four original operational metrics in the comparative analysis.

## 3.3. Behaviour and regulation coupling model

Driving, charging, queueing and resting are represented within one event-driven transition model. A vehicle action may contain a driving phase followed by a non-driving phase, during which queueing, charging and regulatory rest can overlap. Figure 2 provides a conceptual illustration of this coupling.

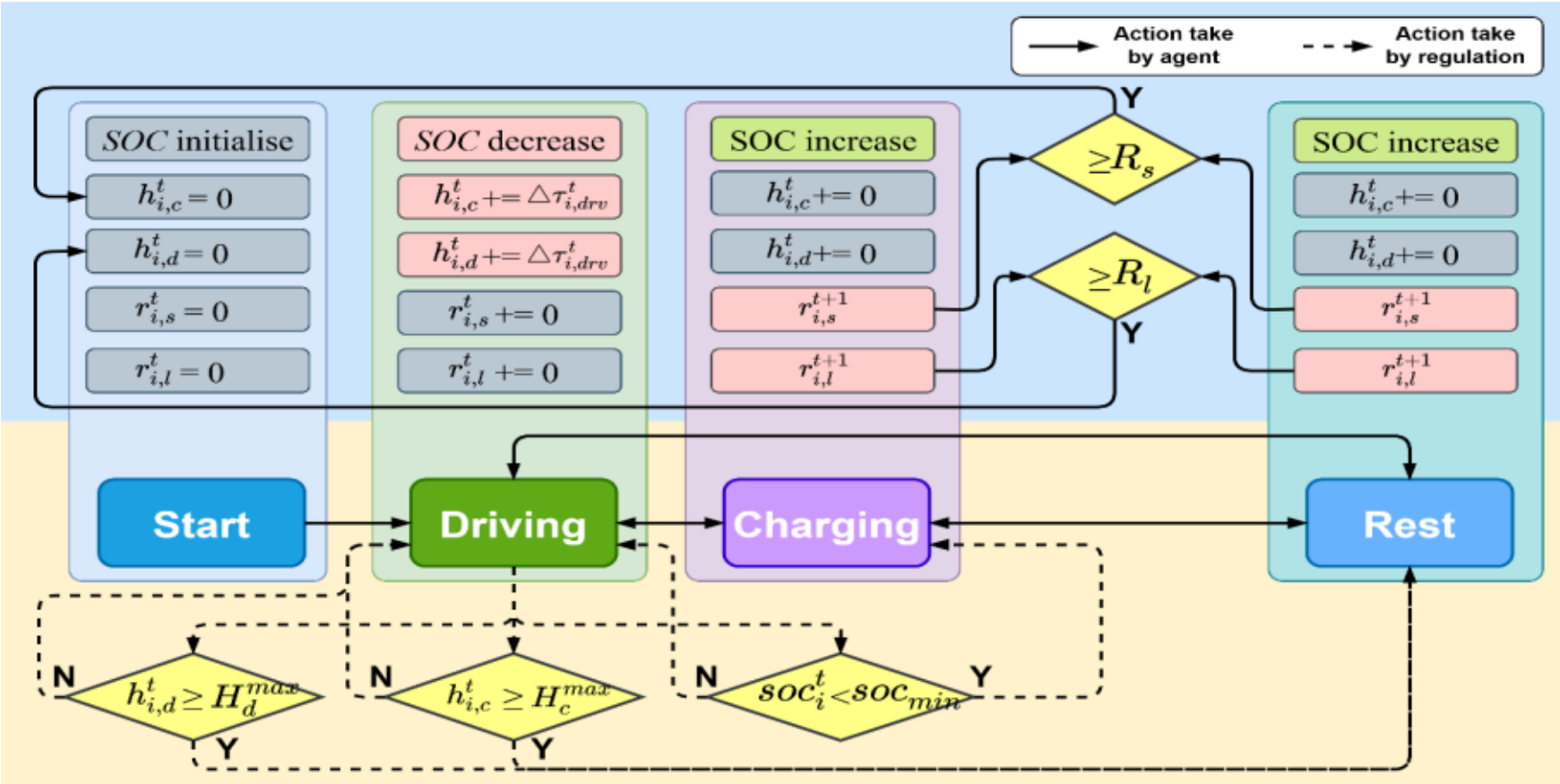

Figure 2. Conceptual charging-rest coupling under driver-hours and energy constraints.

3.3.1. Unified operational dynamics

Each decision transition is resolved through a driving phase and, where selected or required, a non-driving phase. The phases are described below.

**Driving**

After selecting $\text{drive}_k$ or $\text{charge}_k$ at decision step $t$, vehicle $i$ travels from node $p_i^t$ to the selected downstream station $p_i^{t+1}$. The travelled distance is:

$$d_i^t = D(p_i^t, p_i^{t+1}) \tag{8}$$

The corresponding driving time $\Delta\tau_{i,drv}^t$ is:

$$\Delta\tau_{i,drv}^t = \frac{d_i^t}{v} \tag{9}$$

Here, $\mathrm{D}(\cdot)$ is the route-distance function and $v$ is vehicle speed.

Let $\overline{soc}_i^t$ denote the SOC immediately after the driving phase:

$$\overline{soc}_i^t = soc_i^t - \frac{\mu d_i^t}{B} \tag{10}$$

Here, $\mu$ is energy consumption per unit distance and $B$ is battery capacity. For a rest action, no driving occurs and $\overline{soc}_i^t = soc_i^t$.

A driving action also increases the continuous- and daily-driving counters $h_{i,c}^t$ and $h_{i,d}^t$:

$$h_{i,c}^{t+1} = h_{i,c}^t + \Delta\tau_{i,drv}^t \tag{11}$$

$$h_{i,d}^{t+1} = h_{i,d}^t + \Delta\tau_{i,drv}^t \tag{12}$$

A driving phase breaks the preceding uninterrupted non-driving interval. The counters $r_{i,s}^t$ and $r_{i,l}^t$ therefore restart when the next non-driving phase begins.

Non-driving phase (queueing, charging and resting)

The non-driving duration depends on the selected action. Under $\text{charge}_k$, the vehicle waits for the selected charger and then charges. Under rest, the elapsed duration equals the required short- or long-rest period, within which charging may occur.

$$\Delta\tau_{i,\text{non-drv}}^t = \begin{cases} w_i^{t,\rho} + c_i^{t,\rho} & \text{if } charge_k \text{ is selected} \\ R_i{}^t & \text{if } rest \text{ is selected} \end{cases} \tag{13}$$

where $w_{i,\rho}^t$ and $c_{i,\rho}^t$ are the waiting and charging durations. The required rest duration $R_i^t$ is determined by the active driver-hours requirement:

$$R_i{}^t = \begin{cases} R_s & \text{if a short rest is required} \\ R_l & \text{if a long rest is required} \end{cases} \tag{14}$$

Here, $R_\mathrm{s}$ and $R_\mathrm{l}$ are the statutory minimum short- and long-rest durations.

For charger type $\rho$, FIFO waiting time is obtained from the earliest charger-availability time stored in $q^t$:

$$w_i^{t,\rho} = \max(0, q_{p_i{}^{t+1},\rho}^t - (\tau_i^t + \Delta\tau_{i,\text{drv}}^t)) \tag{15}$$

Here, $q_{p_i^{t+1},\rho}^t$ is the earliest availability of charger type $\rho$ at station $p_i^{t+1}$. The charger-type index satisfies:

$$\rho \in \{f, s\} \tag{16}$$

The labels $f$ and $s$ denote fast and slow charging, with powers $P_f$ and $P_s$, respectively.

Queueing and charging are treated as continuous non-driving time. They therefore contribute to a required rest when they form an uninterrupted interval at the same station. A rest action can also use an available charger without extending the prescribed rest duration.

For an uninterrupted same-station non-driving interval, the short- and long-rest counters are updated as:

$$r_{i,s}^{t+1} = r_{i,s}^{t} + \Delta\tau_{i,non-drv}^{t} \tag{17}$$

$$r_{i,l}^{t+1} = r_{i,l}^{t} + \Delta\tau_{i,non-drv}^{t} \tag{18}$$

The counters $r_{i,s}^{t}$ and $r_{i,l}^{t}$ measure the same continuous non-driving interval against different regulatory thresholds. Section 3.3.2 specifies their effects on the driving-time counters.

3.3.2. Regulation constrained state transitions

The driving actions are constrained by energy feasibility, driving and rest regulations:

**Energy feasibility**

Driving actions can only be taken if the minimum SOC constraint is met:

$$soc_i^t - \frac{\mu d_i^t}{B} \geq soc_{min} \tag{19}$$

Here, $soc_{\min}$ is the minimum allowable SOC.

**Driving regulations**

Following the UK assimilated drivers' hours rules for goods vehicles [6], driving and rest actions must satisfy the following constraints:

$$h_{i,c}^t + \Delta\tau_{i,drv}^t \leq H_c^{max} \tag{20}$$

$$h_{i,d}^t + \Delta\tau_{i,drv}^t \leq H_d^{max} \tag{21}$$

Here, $H_c^{max}$ and $H_d^{max}$ are the maximum continuous and daily driving times.

**Rest regulations**

The continuous-driving counter $h_{i,c}^t$ is reset when the short-rest threshold is reached. Both $h_{i,c}^t$ and $h_{i,d}^t$ are reset when the long-rest threshold is reached:

$$h_{i,c}^t = 0, if\ r_{i,s}^{t+1} \geq R_s \tag{22}$$

$$h_{i,c}^t, h_{i,d}^t = 0, if\ r_{i,l}^{t+1} \geq R_l \tag{23}$$

3.3.3. Charging dynamics under shared infrastructure

The pre-charging SOC is $\overline{soc}_i^t$. It equals the post-driving SOC for $\text{charge}_k$ and the current SOC for rest. Purchased energy under charger type $\rho$ is:

$$E_{i,\rho}^t = \begin{cases} \left(soc_{max}^{\rho} - \overline{soc}_i^t\right)B, & \text{if charge}_k \text{ is selected} \\ \min\left\{\left(soc_{max}^{\rho} - \overline{soc}_i^t\right)B, P_\rho \max\left(0, R_i^t - w_{i,\rho}^t\right)\right\}, & \text{if rest is selected} \end{cases} \tag{24}$$

The charger-specific upper SOC $soc_{max}^{\rho}$ is lower for fast charging and equals full charge for slow charging. Under rest, purchased energy is also limited by charger power and the rest time remaining after any admissible wait.

The charging duration $c_{i,\rho}^t$ is:

$$c_i^{t,\rho} = \frac{E_i^{t,\rho}}{P^\rho} \tag{25}$$

The SOC at the next decision event is:

$$soc_i^{t+1} = \overline{soc}_i^t + \frac{E_{i,\rho}^t}{B} \tag{26}$$

The resulting charging expenditure $C_{i,\rho}^t$ is:

$$C_i^{t,\rho} = E_i^{t,\rho} \cdot \kappa^\rho(\tau_i^t + \Delta\tau_{i,\mathrm{drv}}^t + w_i^{t,\rho}) \tag{27}$$

Here, $\kappa_\rho\left(\tau_i^t + \Delta\tau_{i,drv}^t + w_{i,\rho}^t\right)$ is the electricity tariff at charging start time.

3.3.4. Coupled decision mechanism

The operational dynamics above determine the legal action set before policy evaluation. A driving or charging action is removed when the target station cannot be reached above the minimum SOC, when either driving-time limit would be exceeded, or when the required charging service is unavailable. A rest action becomes mandatory when the continuous or daily driving allowance cannot support the next route segment. During a short rest, immediately available fast charging may be used. During a long rest, slow charging is prioritised, with fast charging used when appropriate. A daily-limit-triggered long rest may use queued slow charging when it can be completed within the rest window. The resulting legal action set is:

$$\mathcal{A}_i^t = \{a \in \mathcal{A} | a \text{ is energy-, infrastructure-, and regulation-feasible under } s^t\} \tag{28}$$

The common vocabulary $\mathcal{A}$ and vehicle-specific legal set $\mathcal{A}_i^t$ are defined in Section 3.2.3. The policy assigns probability only to actions executable from state $s^t$.

## 3.4. Real-world data-driven route modelling

Before the MARL decision process, the route-generation layer combines road-network, charging-infrastructure, and scenario-specific disruption data to construct an operational route. Flood-related road closures can remove affected road segments, while power outages can make charging stations unavailable. The resulting route and charging opportunities define the spatial decision context supplied to the environment.

The underlying road network is represented by the undirected weighted multigraph $G_0$:

$$G_0 = (V_0, E_0) \tag{29}$$

Here, $V_0$ and $E_0$ are the road-node and road-link sets. Each edge $(u, v)$ has length $\ell_{uv}$ and road-class label $h(u, v)$.

Let $\mathcal{H}$ contain the retained motorway, motorway-link, trunk, and trunk-link road classes:

$$\mathcal{H} = \{\text{motorway}, \text{motorway_link}, \text{trunk}, \text{trunk_link}\} \tag{30}$$

The HGV-compatible edge set $E_{\mathrm{HGV}}$ is:

$$E_{\mathrm{HGV}} = \{(u, v) \in E_0 \mid h(u, v) \in \mathcal{H}\} \tag{31}$$

The corresponding node set $V_{\mathrm{HGV}}$ contains all endpoints of those edges:

$$V_{\mathrm{HGV}} = \{v \in V_0 \mid \exists (u, w) \in E_{\mathrm{HGV}}: v \in \{u, w\}\} \tag{32}$$

The reduced HGV road network $G_{\mathrm{HGV}}$ is:

$$G_{\mathrm{HGV}} = (V_{\mathrm{HGV}}, E_{\mathrm{HGV}}) \tag{33}$$

Let $\mathcal{U} = \mathcal{U}_{\text{road}} \cup \mathcal{U}_{\text{station}}$ denote active disruptions. The first subset contains road closures, including flood-related closures, and the second contains charging-infrastructure outages. Event $e_k$ is associated with geographic region $\Omega_k$, while $\text{loc}(\cdot)$ returns a road node or station location. The affected road-node and incident-edge sets, $V_U$ and $E_U$, are:

$$V_{\mathcal{U}} = \{v \in V_{\text{HGV}} \mid \exists e_k \in \mathcal{U}_{\text{road}}: loc(v) \in \Omega_k\} \tag{34}$$

$$E_{\mathcal{U}} = \{(u, v) \in E_{\text{HGV}} \mid u \in V_{\mathcal{U}} \ or \ v \in V_{\mathcal{U}}\} \tag{35}$$

Let $\mathcal{S}$ denote the set of $M$ charging stations:

$$\mathcal{S} = \{s_1, s_2, \dots, s_M\} \tag{36}$$

Station $s_j$ has location $(x_j, y_j)$, fast- and slow-charger counts $n_{j,f}$ and $n_{j,s}$, and corresponding powers $P_{j,f}$ and $P_{j,s}$:

$$s_j = (x_j, y_j, n_j{}^f, n_j{}^s, P_j{}^f, P_j{}^s) \tag{37}$$

A station is excluded when it lies in a road-disruption or station-unavailability region. The excluded set is $\mathcal{S}_U$, and the available station set is $\mathcal{S}'$:

$$\mathcal{S}' = \mathcal{S} \setminus \mathcal{S}_{\mathcal{U}} \tag{38}$$

$$\mathcal{S}_{\mathcal{U}} = \left\{s_j \in \mathcal{S} \mid \exists e_k \in (\mathcal{U}_{\text{road}} \cup \mathcal{U}_{\text{station}}): loc(s_j) \in \Omega_k\right\} \tag{39}$$

After removing the affected nodes and edges, the restricted road network $G'$ is:

$$G' = (V', E') \tag{40}$$

$$V' = V_{\text{HGV}} \setminus V_{\mathcal{U}} \tag{41}$$

$$E' = E_{\text{HGV}} \setminus E_{\mathcal{U}} \tag{42}$$

After origin $o$ and destination $d$ are mapped to their nearest available nodes, route $r^*$ is obtained by a distance-weighted shortest-path search on $G'$:

$$r^* = \underset{r \in \mathcal{R}_{o,d}(G')}{\arg\min} \sum_{(u,v) \in r} \ell_{uv} \tag{43}$$

Here, $\mathcal{R}_{o,d}(G')$ is the set of feasible paths from $o$ to $d$, and $\ell_{uv}$ is the length of edge $(u, v)$. The implementation solves Equation (43) using Dijkstra's algorithm.

The route-level charging set $\mathcal{S}_r^*$ retains each available station whose geographical distance from route $r^*$ does not exceed threshold $\delta$. Here, $\text{dist}(s_j, r^*)$ denotes that distance:

$$\mathcal{S}_{r^*} = \left\{s_j \in \mathcal{S}' \mid dist(s_j, r^*) \leq \delta\right\} \tag{44}$$

# 4. Horizon-Preview and Graph-Coordinated PPO

Horizon-Preview and Graph-Coordinated Proximal Policy Optimization (HGC-PPO) converts each vehicle's feasible choices into a prospective action-consequence graph before action selection. At decision event $t$, horizon preview evaluates every legal candidate using the operational model. Candidate consequences are connected to the station, charger-type, and time-slot resources they may use. Two bidirectional GNN message-passing rounds exchange information through these shared resources. A parameter-shared actor scores the coordinated candidates and selects vehicle actions in parallel, while a fleet-level critic supplies the value estimate for PPO training. Figure 3 presents the complete architecture.

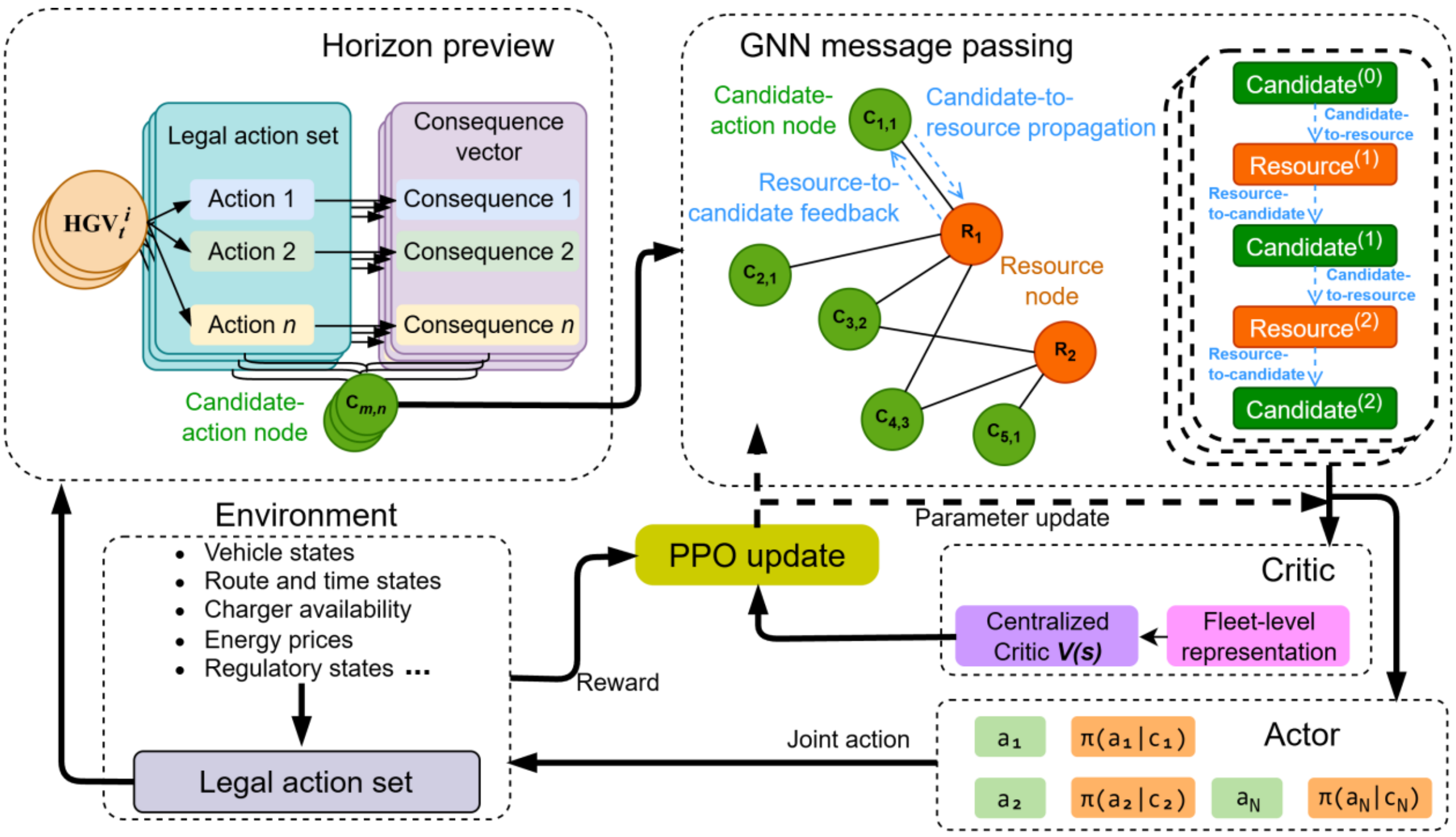

Figure 3. Architecture of HGC-PPO.

## 4.1. Horizon preview and candidate-resource graph construction

For each active vehicle $i$ and legal candidate action $a$ in $\mathcal{A}_i^t$, the action is applied to an isolated copy of fleet state $s^t$. No physical vehicle state, charger queue, or reservation is committed. The immediate calculation returns a consequence vector and the shared resources associated with the candidate. For a travelling action, a fixed maximal-legal-drive scenario is evaluated until the next compulsory service boundary. This extension exposes the next unavoidable rest or energy demand without rolling out the learned policy:

$$\left(\mathbf{x}_{i,a}^t, \Gamma_{i,a}^t\right) = \mathrm{HP}(s^t, i, a), \quad a \in \mathcal{A}_i^t \tag{45}$$

In Equation (45), HP(·) is the deterministic horizon-preview operator. Its outputs are the 16-dimensional consequence vector $x_{i,a}^t$ and prospective resource set $\Gamma_{i,a}^t$. The vector records action type, downstream candidate, projected arrival and elapsed time, queue and charging durations, purchased energy and cost, rest duration, charging-rest overlap, post-action SOC, route progress, forced long-rest status, and arrival status. The resource set contains every station, charger type, and half-hour time slot used by the immediate action or its next compulsory service.

Let $\mathcal{I}^t$ be the active-vehicle subset at decision event $t$. Candidate nodes are indexed by vehicle-action pairs $(i, a)$, whereas resource node $q$ represents one station, charger type, and time slot. An edge is created when a candidate consequence depends on that resource:

$$\begin{aligned} \mathcal{V}_C^t &= \left\{(i, a) \mid i \in \mathcal{I}_t, a \in \mathcal{A}_i^t\right\} \\ \mathcal{V}_R^t &= \bigcup_{(i,a)\in\mathcal{V}_C^t} \Gamma_{i,a}^t \\ \mathcal{E}_{CR}^t &= \left\{((i, a), q) \mid q \in \Gamma_{i,a}^t\right\} \end{aligned} \tag{46}$$

$\mathcal{V}_C^t$, $\mathcal{V}_R^t$, and $\mathcal{E}_{CR}^t$ are the candidate-node, resource-node, and candidate-resource edge sets. Together they define the prospective graph $\mathcal{G}^t = (\mathcal{V}_C^t, \mathcal{V}_R^t, \mathcal{E}_{CR}^t)$. Each resource node contains normalised route location, charger type, time slot, port capacity, available-port share, electricity price, candidate demand, and demand-to-capacity ratio.

## 4.2. Bidirectional GNN message passing

The vehicle observation, candidate consequence, and resource features are encoded into a common hidden dimension. The initial candidate embedding $c_{i,a}^{t,(0)}$ combines the vehicle and candidate encodings, while $r_q^{t,(0)}$ is the initial encoding of resource $q$. Two bidirectional message-passing rounds are then applied using mean aggregation [40] . Let $\ell \in \{1, 2\}$ index the round. Each round first sends candidate information to the connected resources:

$$\bar{c}_q^{t,(\ell)} = \frac{\sum_{(i,a)\in\mathcal{N}(q)} c_{i,a}^{t,(\ell-1)}}{|\mathcal{N}(q)|}$$
$$r_q^{t,(\ell)} = \tanh\left(W_R^{(\ell)}[r_q^{t,(\ell-1)} \parallel \bar{c}_q^{t,(\ell)}] + b_R^{(\ell)}\right) \tag{47}$$

$\mathcal{N}(q)$ is the set of candidate nodes connected to resource $q$; $\bar{c}_q^{t,(\ell)}$ is their mean candidate message; and $r_q^{t,(\ell)}$ is the updated resource embedding. The matrices $W_R^{(\ell)}$ and biases $b_R^{(\ell)}$ are trainable resource-update parameters. The updated resource information is then returned to every connected candidate:

$$\bar{r}_{i,a}^{t,(\ell)} = \frac{\sum_{q\in\mathcal{N}(i,a)} r_q^{t,(\ell)}}{|\mathcal{N}(i,a)|}$$
$$c_{i,a}^{t,(\ell)} = \tanh\left(W_C^{(\ell)}[c_{i,a}^{t,(\ell-1)} \parallel \bar{r}_{i,a}^{t,(\ell)}] + b_C^{(\ell)}\right) \tag{48}$$

$\mathcal{N}(i,a)$ is the resource neighbourhood of candidate $(i,a)$, and $\bar{r}_{i,a}^{t,(\ell)}$ is its mean resource feedback. The matrices $W_C^{(\ell)}$ and biases $b_C^{(\ell)}$ update candidate embedding $c_{i,a}^{t,(\ell)}$. The symbol $\parallel$ denotes concatenation and tanh is the activation function. After round $\ell = 2$, each candidate contains its own consequence and a learned summary of other candidates connected through the same resources.

## 4.3. Shared actor and parallel multi-agent policy

A parameter-shared actor maps each coordinated candidate embedding $c_{i,a}^{t,(2)}$ to logit $z_{i,a}^t$. The legal-action mask from Section 3.3.4 is applied before softmax normalisation:

$$z_{i,a}^t = \mathbf{w}^T \mathbf{c}_{i,a}^{t,(2)} + b_\pi$$
$$\pi_\theta(a \mid i,t) = \frac{m_{i,a}^t \exp(z_{i,a}^t)}{\sum_{b\in\mathcal{A}_i^t} m_{i,b}^t \exp(z_{i,b}^t)} \tag{49}$$

where $w$ and $b_\pi$ are the shared actor-head parameters, $m_{i,a}^t$ equals one for a legal action and zero otherwise, and $\pi_\theta$ is the policy with parameters $\theta$. Because graph coordination has already modified every candidate embedding, the joint action can be sampled without sequential action conditioning:

$$\pi_\theta(a^t \mid s^t, \mathcal{G}^t) = \prod_{i\in\mathcal{I}_t} \pi_\theta(a_i{}^t \mid i,t) \tag{50}$$

Equation (50) factorises the joint policy over active vehicles $\mathcal{I}^t$. Here, $a^t$ is the joint fleet action and $\mathcal{G}^t$ is the prospective graph defined after Equation (46). Actions are produced in parallel, while their logits remain coupled through graph message passing.

## 4.4. Fleet-level critic and PPO optimisation

The critic evaluates the fleet rather than an individual vehicle. For each active vehicle, its legal coordinated candidates are averaged into a vehicle-level candidate context. These contexts are pooled across the fleet and concatenated with the mean vehicle embedding:

$$\mathbf{c}_i{}^t = \frac{\sum_{a\in\mathcal{A}_i^t} \mathbf{c}_{i,a}^{t,(2)}}{|\mathcal{A}_i^t|}$$
$$\mathbf{u}^t = [\frac{\sum_{i\in\mathcal{I}_t} \mathbf{c}_i{}^t}{|\mathcal{I}_t|} \parallel \frac{\sum_{i\in\mathcal{I}_t} \mathbf{v}_i{}^t}{|\mathcal{I}_t|}] \tag{51}$$
$$V_\psi(s^t) = f_V\big(\text{stopgrad}(\mathbf{u}^t)\big)$$

where $c_i^t$ is the mean coordinated-candidate context for vehicle $i$, $v_i^t$ is its encoded observation, and $u^t$ is the fleet-level representation. Critic $V_\psi$ has parameters $\psi$ and is implemented by network $f_V$. The operator $\text{stopgrad}(\cdot)$ detaches the pooled representation, so value loss updates the critic without changing the shared encoders. Generalised advantage estimation converts shared reward $r_t$ from Section 4.5 into a fleet-level advantage [41]:

$$\delta_t = r_t + \gamma V_\psi(s^{t+1}) - V_\psi(s^t)$$
$$\hat{A}_t = \sum_{k=0}^{T-t-1} (\gamma\lambda)^k \delta_{t+k}, \quad \hat{R}_t = \hat{A}_t + V_\psi(s^t) \tag{52}$$

$\delta_t$ is the temporal-difference residual, $\gamma$ is the discount factor, $\lambda$ is the GAE parameter, and $T$ is the rollout horizon. The quantities $\hat{A}_t$ and $\hat{R}_t$ are the fleet-level advantage and value target. This shared advantage is assigned to each active vehicle's selected action. The actor is trained with the clipped PPO surrogate [38] , and the complete loss includes value error and policy entropy:

$$\rho_{i,t}(\theta) = \frac{\pi_\theta\big(a_i^t \mid i,t\big)}{\pi_{\theta old}\big(a_i^t \mid i,t\big)}$$
$$\mathcal{L}_{\text{clip}} = \mathrm{E}_t\left(\sum_{i\in\mathcal{I}^t} \min\big(\rho_{i,t}\hat{A}_t, \text{clip}\big(\rho_{i,t}, 1-\varepsilon, 1+\varepsilon\big)\hat{A}_t\big)\right) \tag{53}$$
$$\mathcal{L} = -\mathcal{L}_{\text{clip}} + \frac{c_v}{2}\mathrm{E}_t\left(\big(V_\psi(s^t) - \hat{R}_t\big)^2\right) - c_e \mathrm{E}_t\big(\mathcal{H}(\pi_\theta)\big)$$

$\rho_{i,t}(\theta)$ is the likelihood ratio between current policy $\pi_\theta$ and stored behaviour policy $\pi_{\theta old}$. The operator $\text{clip}(\cdot)$ limits this ratio to $[1-\varepsilon, 1+\varepsilon]$, where $\varepsilon$ is the clipping radius. The expectation $\mathrm{E}_t$ averages over rollout decision events. The coefficients $c_v$ and $c_e$ weight the value and entropy terms, $\mathcal{H}(\pi_\theta)$ is policy entropy, and $\mathcal{L}$ is the total minimised loss. PPO updates the encoders, both GNN rounds, the actor, and the critic. Horizon preview and legal-action construction have no trainable parameters.

## 4.5. Shared reward

All vehicles receive the same fleet reward. The legal action set enforces feasibility, while the reward represents operational efficiency. It combines normalised route progress, elapsed and resource-blocked time, charging expenditure, queueing, fleet completion, charging frequency, and high-SOC herd charging:

$$
\begin{gathered}
{r_{\mathrm{op}}}^{t} = \frac{1}{N}\left[w_d \frac{\Delta d_t}{d_0} - w_\tau \frac{\Delta\tau_t + \Delta b_t}{\tau_0} - w_C \frac{\Delta C_t}{C_0} - w_Q \frac{\Delta Q_t}{Q_0}\right] + 1_{\mathrm{done}} \frac{B}{B_0} \\
r_t = \alpha_{\mathrm{team}}\left[{r_{\mathrm{op}}}^{t} - w_N \frac{{\Delta N_{\mathrm{chg}}}^{t}}{N}\right] - \alpha_{\mathrm{coord}} w_H H_t
\end{gathered}
\tag{54}
$$

Equation (54) first forms the normalised operational score $r_{op}^{t}$ and then the final shared reward $r_t$. The increments $\Delta d_t$, $\Delta\tau_t$, $\Delta b_t$, $\Delta C_t$, and $\Delta Q_t$ denote fleet route progress, elapsed time, resource-blocked time, charging expenditure, and queueing time. The term $\Delta N_{chg}^{t}$ counts charging events. The constants $d_0$, $\tau_0$, $C_0$, $Q_0$, and $B_0$ are fixed normalisation scales. Indicator $1_{\mathrm{done}}$ activates fleet-completion bonus $B$.

The coefficients $w_d$, $w_\tau$, $w_C$, $w_Q$, and $w_N$ weight progress, time, charging expenditure, queueing, and charging frequency. The term $H_t$ is the proportion of active vehicles initiating charging above the fixed SOC threshold, weighted by $w_H$. All RL algorithms use the same reward.

# 5. Case study and experimental design

The case study was designed to separate training-task performance, generalisation, and module contribution. All methods were evaluated in the same event-driven environment with common vehicle physics, charging resources, driver-hours rules, legal-action constraints, and reward definition. Formal training was conducted on one demanding long-haul route, whereas transfer was evaluated without additional learning on six unseen routes and four active fleet sizes.

## 5.1. Case-study data and route design

The UK road network, charging-station locations, charger capacities, and time-varying electricity prices were processed through the route-generation framework described in Section 3.4. Each route file records charging station coordinates, cumulative distance, and the numbers of 350kW and 22kW chargers available at each station.

The formal route groups are shown as follows:

1). Training task: Folkestone–Glasgow (794.00 km) with 20 vehicles and 32 en-route charging stations. This route requires repeated charging, short and long rests, and both fast and slow charging.

2). Short generalisation routes: London–Birmingham (162.83 km) and Birmingham–Leeds (148.50 km). These routes can be completed without en-route charging under the configured vehicle specification.

3). Medium generalisation routes: London–Edinburgh (534.33 km) and Tilbury–Glasgow (576.70 km). These routes require intermediate fast charging and short-rest decisions but do not require long rests.

4). Long generalisation routes: Plymouth–Aberdeen (765.88 km) and Brighton–Inverness (787.10 km). These routes require repeated charging and include both short- and long-rest decisions.

5). Fleet-load transfer: every generalisation route was evaluated with fleet number of 5, 10, 15, and 20.

## 5.2. Vehicle, charging and regulatory settings

The vehicle specification represents a long-haul battery-electric HGV. The simulated vehicle uses a 520-kWh battery [42], [43]. Energy consumption is fixed at 1.15 kWh/km and cruising speed at 82 km/h [43], [44]. The initial SOC is 1.0, and the minimum admissible SOC is 0.1. Fast charging terminates at SOC 0.9, whereas slow charging may continue to SOC 1.0 [45].

Driver-hours limits follow the standard UK assimilated rules for goods vehicles [6]. A short rest of 45 min is required after at most 4.5h of continuous driving, and an 11h long rest follows the 9 h daily driving limit. Charging time contributes to an eligible rest period, while access to a fully occupied charger creates a first in, first out (FIFO) waiting order. The operational values used in all formal experiments are summarised in Table 1.

**Table 1. Vehicle, charging and driver-hours parameters used in the formal experiments.**

| *Category* | *Parameter* | *Value* |
|---|---|---|
| Vehicle | Battery capacity | 520 kWh |
| | Initial SOC | 1.00 |
| | Minimum SOC | 0.10 |
| | Energy consumption | 1.15 kWh/km |
| | Cruising speed | 82 km/h |
| Charging | Fast-charger power | 350 kW |
| | Slow-charger power | 22 kW |
| | Fast-charge target SOC | 0.90 |
| | Slow-charge target SOC | 1.00 |
| | Fast-charging price | £1.00/kWh |
| | Slow-charging price | £0.148–£0.361/kWh at 30-min resolution |
| | Queue discipline | FIFO |
| Driver hours | Maximum continuous driving | 4.5 h |
| | Short-rest duration | 0.75 h |
| | Maximum daily driving | 9 h |
| | Long-rest duration | 11 h |

## 5.3. Compared learning and optimisation methods

The proposed HGC-PPO was compared with RL family: PPO, IPPO, MAPPO, and HAPPO. All five RL methods used the same operational environment, feasible-action mechanism, reward, training budget, evaluation schedule, and core PPO hyperparameters.

A deterministic rule-based baseline was included as an operational reference. Vehicles continue driving while energy and driver-hours constraints permit, select charging when additional energy is required, and take mandatory rests when the corresponding driving limit is reached. Its decisions depend on local feasibility rules rather than a learned fleet policy.

Two optimisation Mixed-Integer Linear Programming (MILP) references were recomputed for the 20-vehicle Folkestone–Glasgow training task. MILP-L provides optimistic lower bounds for fleet completion time and charging cost under relaxed shared-resource assumptions. These bounds do not form one executable fleet schedule, so queue time and charging-event frequency are not reported for MILP-L. MILP-U selects among a restricted set of feasible action profiles and fleet assignments, after which the complete schedule is replayed in the same event-driven environment. It is therefore a feasible restricted-profile reference rather than the unrestricted global optimum. The complete formulation and computational settings are provided in Section S2 of the Supplementary Material.

## 5.4. Formal training, generalisation and ablation protocols

Each RL algorithm was trained from an initialisation with seeds 11, 12, and 13 for 20,000 environment transitions per run. No checkpoint or optimiser state was warm-started. Deterministic evaluation was performed every 256 transitions on the training task. Generalisation was evaluated with the frozen best checkpoints from the formal training runs.

The ablation study used a complete 2×2 design over resource-graph message passing and horizon preview. The full HGC-PPO cell reused the formal training runs, and the without-graph, without-horizon, and joint-removal variants were trained from fresh initialisations with the same three seeds and 20,000-transition budget. Network capacity, reward, environment, PPO settings, and checkpoint selection remained the same. Table 2 summarises the formal protocol. Reward coefficients, training settings, implementation details and fixed-policy evaluation times are provided in Section S1 of the Supplementary Material.

**Table 2. Formal training, generalisation, ablation and optimisation-reference protocol.**

| ***Experiment*** | ***Methods or configurations*** | ***Seeds*** | ***Route and fleet*** | ***Budget or evaluation*** |
|---|---|---|---|---|
| Training comparison | HGC-PPO, PPO, IPPO, MAPPO, HAPPO | 11, 12, 13 | Folkestone–Glasgow; 20 vehicles | 20,000 transitions per learning run |
| Generalisation | Five frozen RL policies and Rule-based | 11, 12, 13 for RL methods | Six unseen routes; 5, 10, 15, and 20 active vehicles | No further training |
| Module ablation | Full, without graph, without horizon, and without graph or horizon | 11, 12, 13 | Folkestone–Glasgow; 20 vehicles | 20,000 transitions per new run; full model reused |
| Optimisation references | MILP-L and MILP-U | Not applicable | Folkestone–Glasgow; 20 vehicles | Recomputed once in the matched formal environment |

## 5.5. Performance metrics and reporting

Performance was assessed using four operational metrics:

1). Fleet completion time (h): the elapsed time until the final active vehicle reaches the destination, including driving, charging, resource-blocked queueing, and mandatory rest.

2). Charging cost per vehicle (£/vehicle): the fleet electricity expenditure divided by the number of active vehicles.

3). Queue time per vehicle (h/vehicle): the charger-access waiting time recorded by the environment, averaged over vehicles.

4). Charging events per vehicle (events/vehicle): the mean number of charging sessions completed by each active vehicle.

For RL methods, every training-task, generalisation, and ablation result is reported as the arithmetic mean and sample standard deviation across seeds 11, 12, and 13. Within-seed differences are retained for paired directional interpretation. Rule-based and MILP results are deterministic references and are therefore reported without artificial seed variance.

For the generalisation analysis, condition-level metrics were first averaged across the three training seeds. HGC-PPO was then compared with each RL baseline over the 24 unseen route–fleet conditions using two-sided Wilcoxon signed-rank tests. Holm correction was applied across the four baseline comparisons separately for each metric. Mean paired differences and 95% percentile bootstrap confidence intervals were estimated from 50,000 resamples of the matched conditions.

# 6. Results

The evaluation examined performance on the training task, generalisation across routes and fleet loads, the contribution of horizon preview and graph message passing, and charging behaviour on the long route. Fleet completion time, charging cost, queueing time and charging events were analysed together.

## 6.1. Training-task performance

Figure 4 compares the formal results on the 794km Folkestone-Glasgow route with 20 vehicles. HGC-PPO achieved the lowest mean charging cost at £504.32 per vehicle, while maintaining a fleet completion time of 24.920h, queueing of 7.167h per vehicle and 2.867 charging events per vehicle. This combination gave HGC-PPO the strongest overall operating balance among the RL methods. The exact seed-level values underlying Figure 4 are reported in Table C1 of the Supplementary Material.

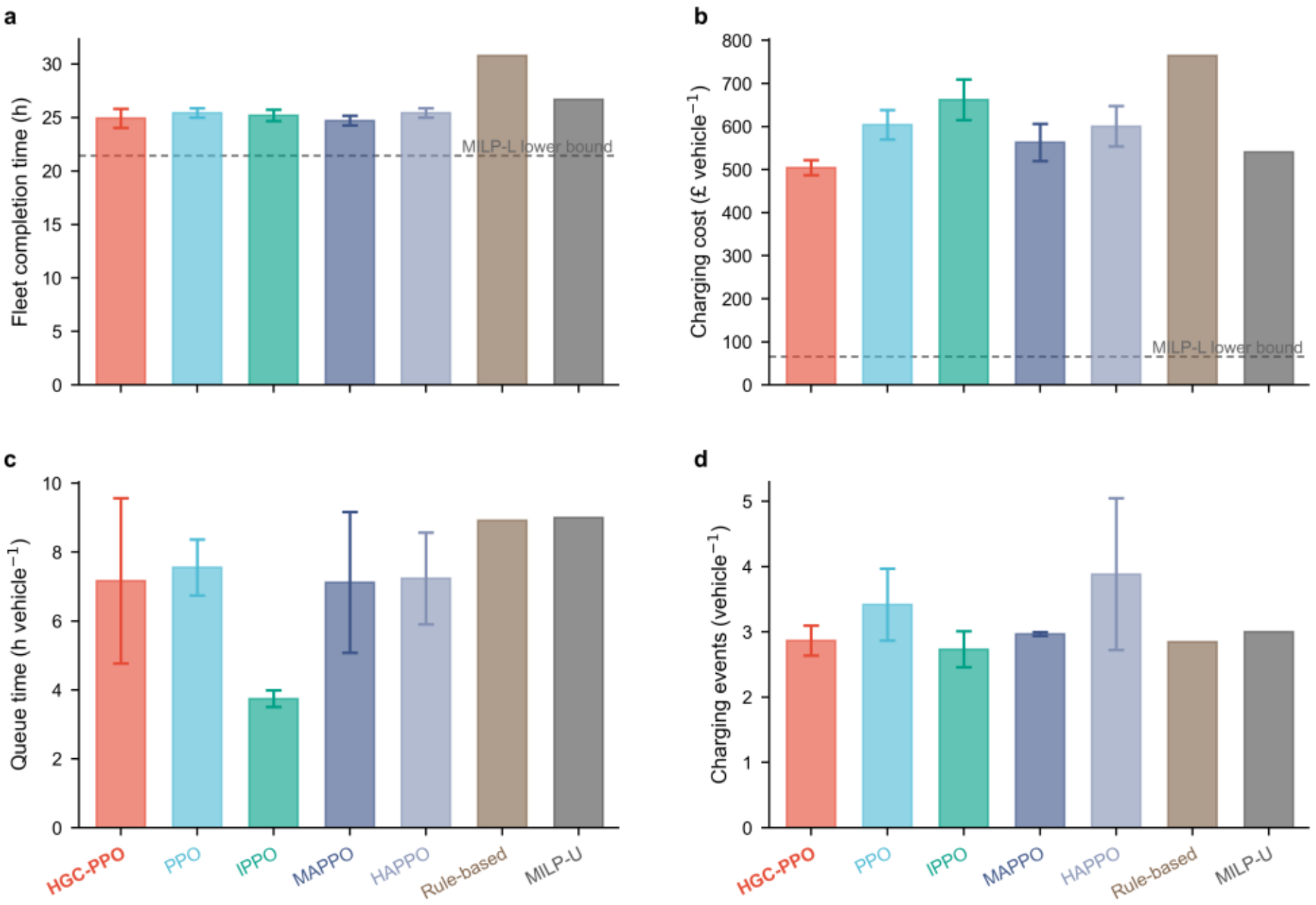

Figure 4. Formal training-task performance.

MAPPO completed the route 0.220h earlier than HGC-PPO, but cost £58.74 more per vehicle, with comparable queueing and slightly more charging events. IPPO achieved the lowest queueing time and the fewest charging events but completed 0.279h later than HGC-PPO and cost £157.83 more per vehicle. Its queueing advantage therefore did not translate into faster fleet operation and was accompanied by the highest charging cost among the RL policies. PPO and HAPPO were slower, more expensive and charged more frequently than HGC-PPO.

Charging frequency provides an additional measure of policy quality. Rule-based and MILP-U produced 2.850 and 3.000 charging events per vehicle, respectively, providing independent references for charging pattern. HGC-PPO reproduced it with 2.867 ± 0.231 events. PPO and HAPPO increased the frequency to 3.417 and 3.883 events without reducing fleet completion time or charging cost, indicating more fragmented charging rather than an efficiency gain.

MILP-L provided separate idealised lower bounds of 21.433 h and £65.87 per vehicle for completion time and charging cost. The executable MILP-U reference returned 26.683h of completion time,

£540.83 per vehicle, 9h of queueing per vehicle and 3.000 charging events per vehicle. HGC-PPO completed 1.763h earlier and cost £36.51 less per vehicle than MILP-U, while reducing queueing by 1.833h and charging frequency by 0.133 events per vehicle.

## 6.2. Generalisation across routes and fleet loads

Figure 5 compares the five RL methods on six unseen routes with 5, 10, 15 and 20 active vehicles.

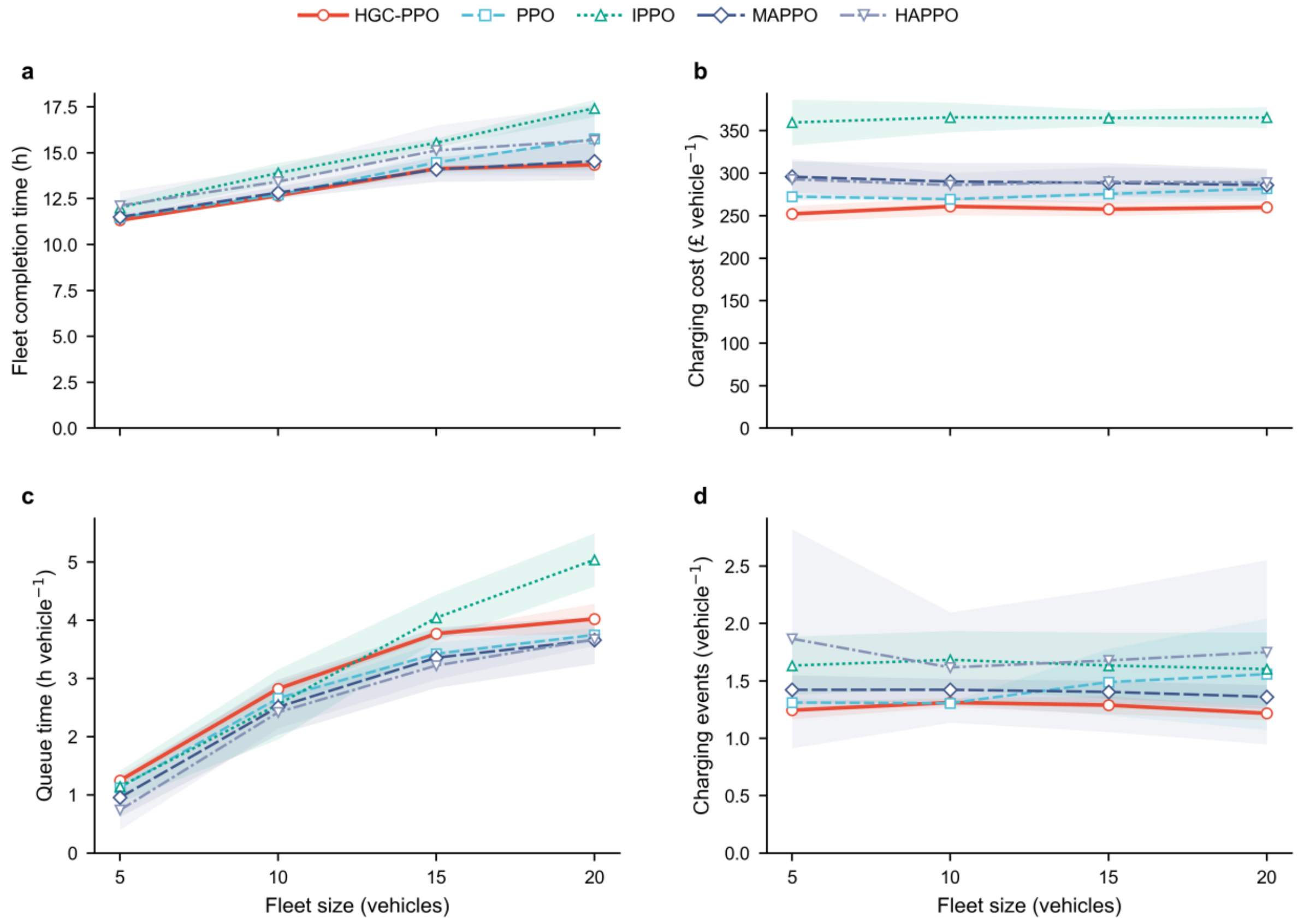

Figure 5. Generalisation of the RL methods.

Across all 24 route-fleet conditions, HGC-PPO achieved the lowest average fleet completion time at 13.116h and the lowest charging cost at £257.58 per vehicle. It also recorded the fewest charging events among the RL methods at 1.265 per vehicle. Figure 6 reports paired differences defined as baseline minus HGC-PPO, such that positive values favour HGC-PPO, with horizontal bars denoting 95% bootstrap confidence intervals. The charging-cost and charging-frequency intervals remained above zero relative to every RL baseline, with Holm-adjusted paired tests confirming both advantages (all adjusted $p \leq 0.0028$). Completion-time advantages were also confirmed relative to PPO, IPPO and HAPPO (all adjusted $p \leq 0.019$), while HGC-PPO retained the lowest overall mean completion time across all methods. Queue time was treated as a diagnostic of shared-resource interaction and interpreted together with fleet completion time, because its operational consequence is reflected in when the fleet completes the journey. Overall, the paired analysis shows that HGC-PPO converted prospective resource coordination into lower charging cost, fewer charging events and efficient fleet completion across unseen operating conditions.

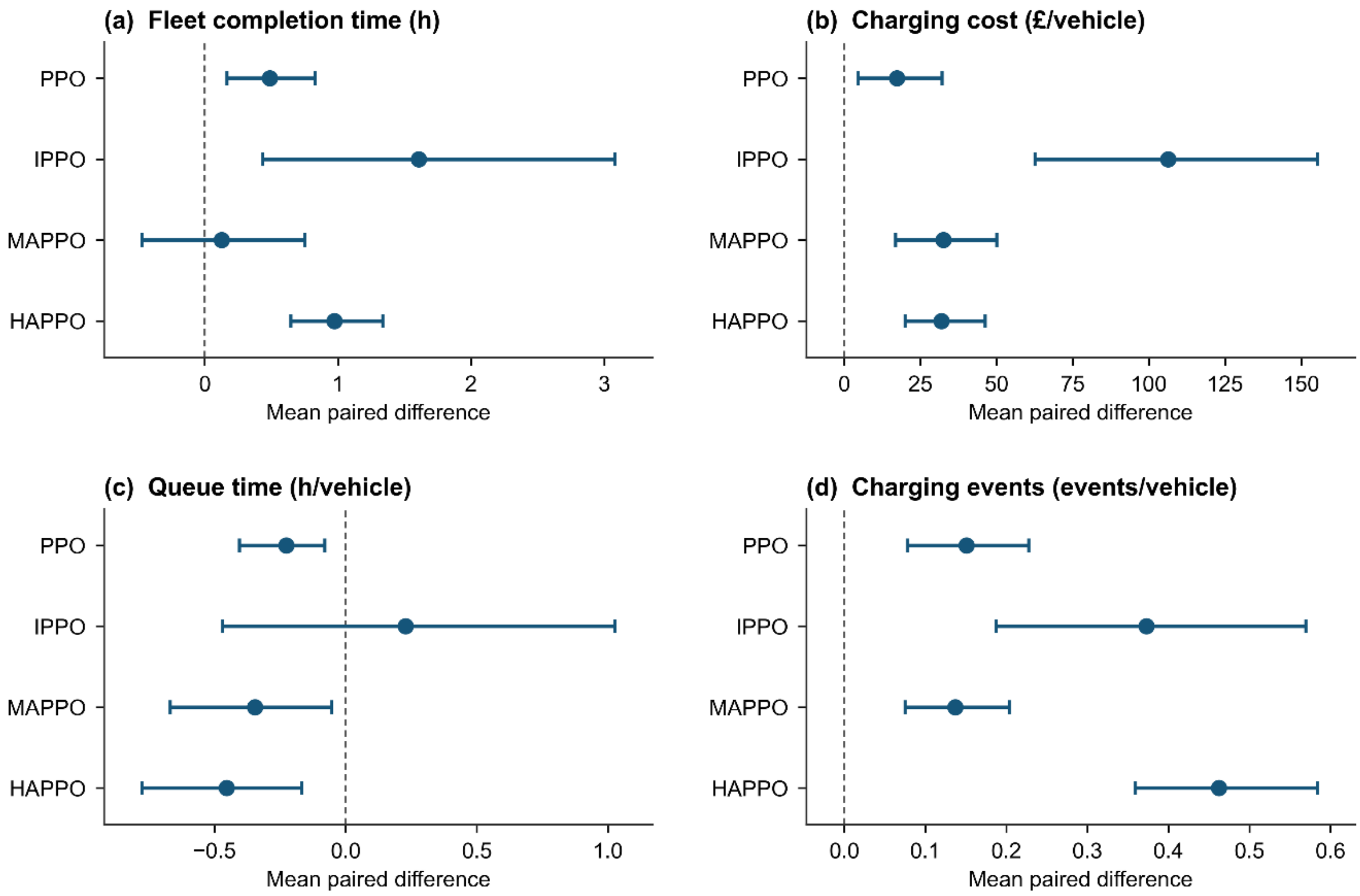

Figure 6. Mean paired differences between the RL baselines and HGC-PPO.

The same pattern was observed across route classes. HGC-PPO avoided unnecessary charging on the short routes and used one charging event per vehicle on the medium routes. On the long routes, it retained the lowest charging cost and the fewest charging events among the RL methods. Queueing increased under the most demanding long-route conditions, but this did not remove HGC-PPO's cost and charging-frequency advantages. Overall and route-specific results underlying Figure 5 are reported in Tables C2 and C3 of the Supplementary Material.

## 6.3. Ablation study

Figure 7 evaluates the complete prospective coordination architecture through a $2 \times 2$ ablation of horizon preview and candidate-resource graph message passing. The full HGC-PPO architecture achieved the lowest mean fleet completion time, charging cost, queueing time and charging-event frequency among the four configurations. Relative to the joint-removal configuration, it reduced fleet completion time by 6.61%, charging cost by 6.75%, queueing time by 18.71% and charging events by 3.91%.

The ablation reveals the functional dependence between the two stages of HGC-PPO. Horizon preview constructs candidate-action nodes containing the projected operational consequences of each legal choice, while graph message passing coordinates these candidates through the resources they may jointly use. Adding graph coordination to the horizon-preview representation reduced fleet completion time from 25.608 to 24.920 h, charging cost from £516.50 to £504.32 per vehicle, queueing from 7.841 to 7.167 h per vehicle and charging frequency from 2.900 to 2.867 events per vehicle. Graph coordination without horizon preview improved completion time, queueing and charging frequency relative to the joint-removal configuration, but did not reduce charging cost. The complete architecture produced the best four-metric outcome because graph propagation operated on candidate nodes enriched with prospective operational consequences. The seed-level results are reported in Table C4 of the Supplementary Information.

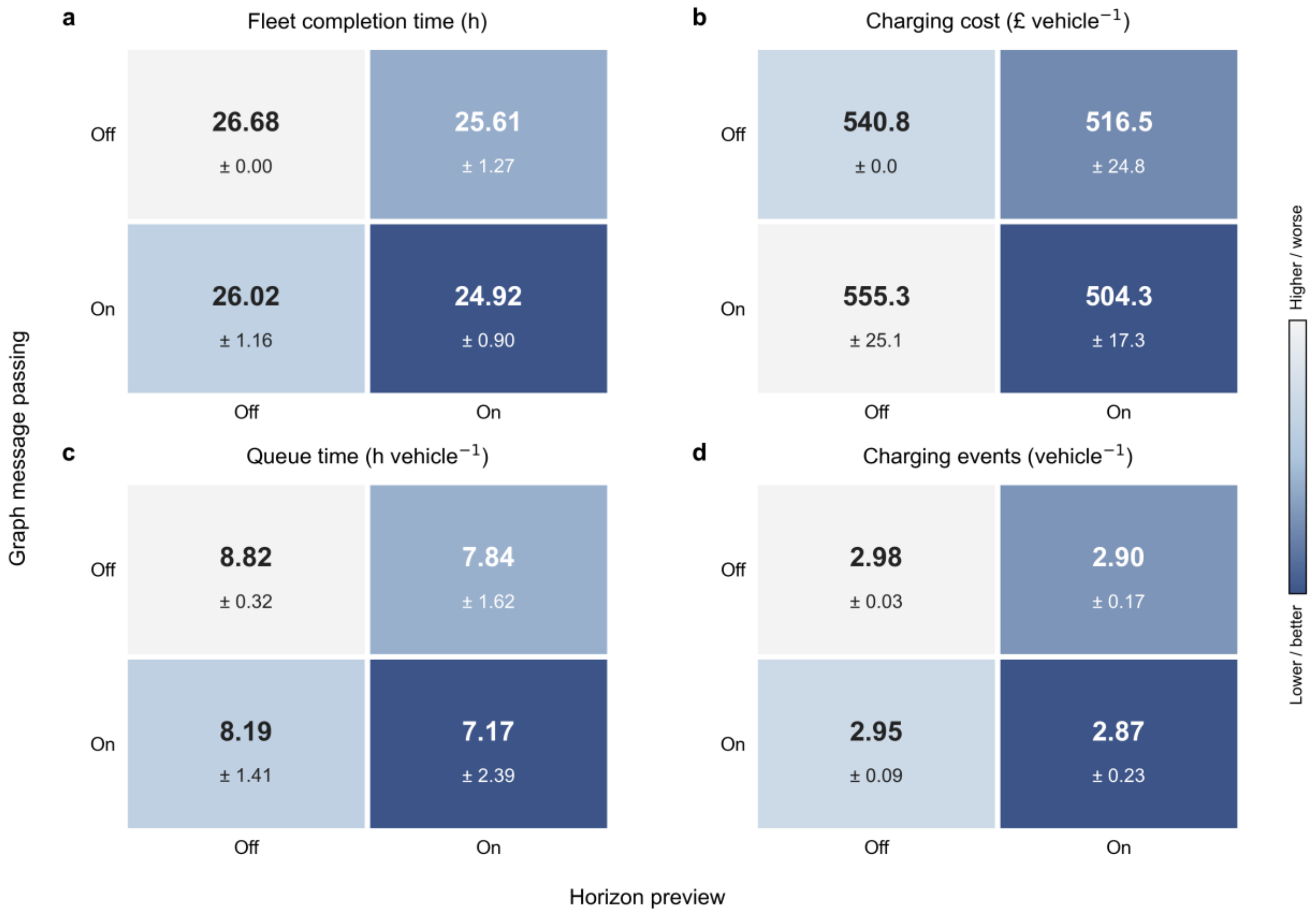

Figure 7. Ablation of graph message passing and horizon preview.

## 6.4. Charging behaviour and resource coordination

Figures 8–10 examine how the methods used charging infrastructure on the Folkestone-Glasgow route with 20 vehicles. The resource-pressure heatmaps in Figure 8 show a common slow-charging bottleneck at station GS10118 during the long-rest period. HGC-PPO produced a less severe and more localised overload pattern at this bottleneck than PPO, MAPPO, HAPPO, Rule-based and MILP-U.

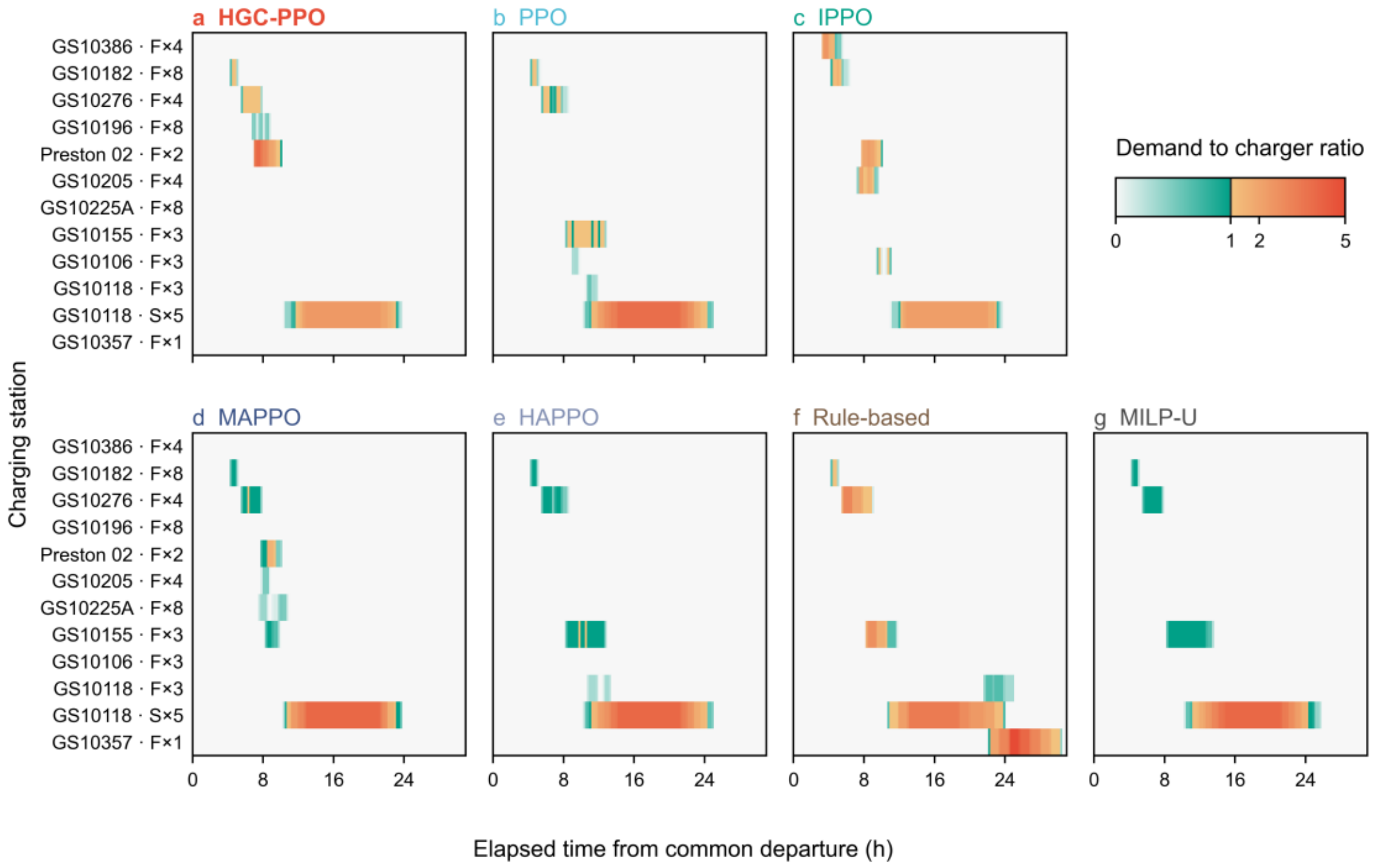

Figure 8. Charging-resource pressure heatmap.

Figure 9 shows that HGC-PPO purchased 553 kWh per vehicle, the lowest value among the seven methods. MILP-U purchased 606 kWh per vehicle, the other RL methods purchased 624–659 kWh, and Rule-based purchased 819 kWh. HGC-PPO also had the lowest energy above the safe terminal reserve, at 108 kWh per vehicle. Its cost advantage therefore resulted from limiting unnecessary energy purchases while completing the task within the battery constraints.

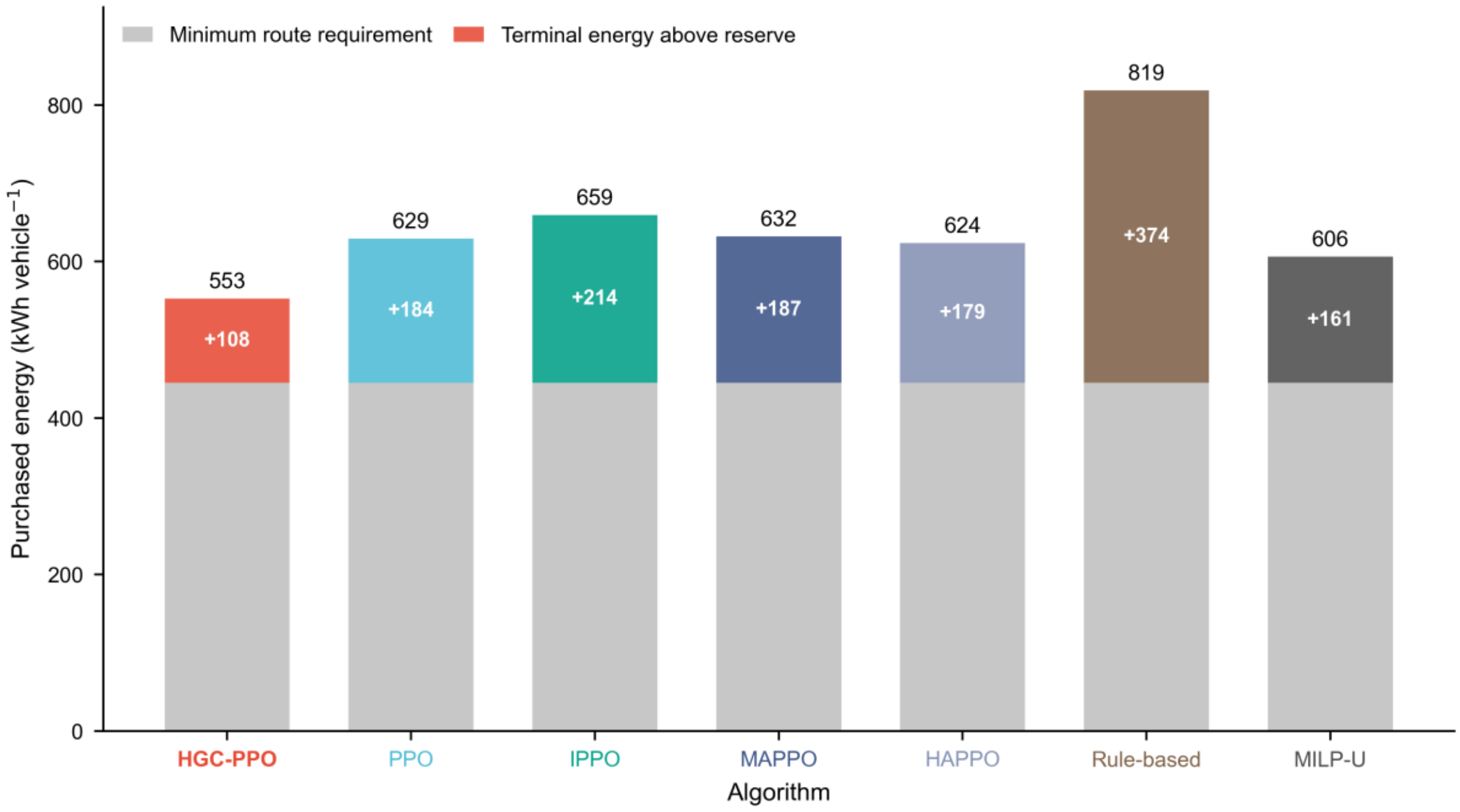

Figure 9. Comparison of purchased energy per vehicle.

Figure 10 further shows that HGC-PPO distributed its charging events across five stations. Individual station shares ranged from 15% to 23%, and the largest share was 23%, jointly the lowest concentration with IPPO. This distribution prevented one station from dominating the charging schedule and represents a practical use of the available fast- and slow-charging resources.

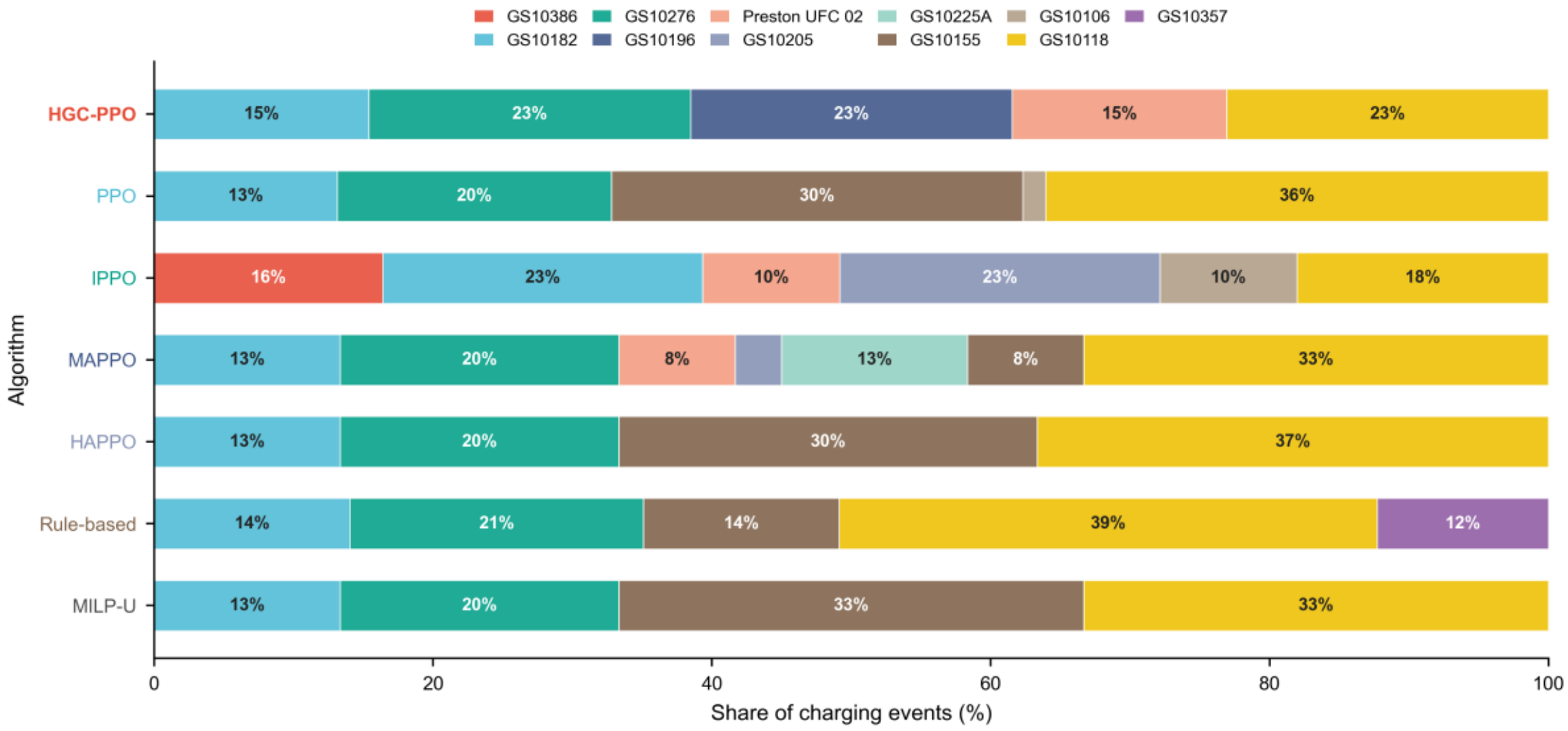

Figure 10. Comparison of charging distribution.

Taken together, the figures show that HGC-PPO combined the lowest purchased energy with distributed station use and reduced persistent pressure at the main slow-charging bottleneck. The method did not minimise every queue-related measure, but it converted charging coordination into lower energy use, limited charging fragmentation and efficient fleet completion.

# 7. Discussion

HGC-PPO's operating advantage can be understood as an information advantage at action selection. All RL methods were trained with the same environment, legal-action mechanism, reward and training budget, but the baseline policies evaluated actions without explicitly constructing their action-conditioned operational futures and prospective resource interactions. HGC-PPO extended this decision context in three directions. It represented the consequences of each legal action rather than only the current vehicle state; projected the vehicle towards the next compulsory service boundary created by its energy and driver-hours conditions; and incorporated the prospective resource demand of other vehicles into each candidate representation. The actor therefore evaluated not only what a vehicle could do immediately, but where the action would lead, which charging or resting obligation it would create, and how that choice would interact with the fleet's use of shared infrastructure.

These information advantages were reflected in the resulting fleet behaviour. HGC-PPO limited unnecessary energy purchases, reduced charging fragmentation and distributed charging events across five stations while maintaining efficient fleet completion. It also produced a more localised pressure pattern at the principal slow-charging bottleneck. The ablation results connected these behaviours to the complete prospective coordination architecture: horizon preview supplied action-conditioned future information, while graph propagation incorporated fleet-wide resource interactions into those action representations. Their integration allowed the RL policy to coordinate driving, charging and resting decisions before their shared-infrastructure consequences were realised.

Centralised optimisation, including the intercity charging-infrastructure model of Bao and Xie [28], resolves resource interaction through a jointly constructed solution. Existing MARL and graph-based charging methods coordinate through critics, rewards, attention mechanisms or graphs of physical entities [8], [20], [21], [39]. HGC-PPO adds an action-level prospective representation: other vehicles' dependence on a charging resource can alter each legal action before the Actor evaluates it, while the policy update remains PPO.

The evidence covers six unseen UK routes and fleet loads of 5-20 vehicles. Queueing growth on the most demanding long routes remains the main operating boundary. Future extensions could couple prospective fleet coordination with renewable-energy-aware charging, allowing charging decisions to respond to the time-varying availability and operational optimisation of renewable generation [46].

# 8. Conclusions

This study developed HGC-PPO, a regulation-aware multi-agent reinforcement learning framework for coordinated eHGV operation under shared charging infrastructure. The framework couples vehicle driving, charging, queueing, resting, SOC and driver-hours compliance within one decision environment, allowing operationally feasible fleet decisions to be learned directly.

Within HGC-PPO, this coordination is implemented through two complementary components. First, horizon preview converts each legal action into a candidate-action node that carries its short-horizon operational consequence. Second, bidirectional GNN message passing updates these candidates through the charging resources shared by different vehicles, so that resource dependencies are represented before an action is selected. The coordinated candidate representations are then evaluated by a shared Actor and trained with PPO. This design makes the consequences of prospective charging choices explicit, instead of leaving the policy to infer them only from delayed reward feedback.

The formal experiments provide consistent evidence for this design. Across three training seeds on the 794-km Folkestone-Glasgow task with 20 vehicles, HGC-PPO achieved the lowest mean charging cost among the RL methods at £504.32 per vehicle, together with a fleet completion time of 24.920h and 2.867 charging events per vehicle. It also completed the task 1.763h earlier and £36.51 per vehicle cheaper than the executable MILP-U reference. Across 24 combinations of six unseen routes and fleet loads from 5 to 20 vehicles, HGC-PPO again obtained the lowest mean fleet completion time, charging cost and charging frequency among the learning-based RL methods: 13.116h, £257.58 per vehicle and 1.265 events per vehicle. The full ablation model outperformed the variant without both proposed modules by 6.61% in fleet completion time, 6.75% in charging cost, 18.71% in queueing time and 3.91% in charging frequency.

The resulting policy also exhibited healthy charging behaviour. On the long training route, HGC-PPO purchased 553 kWh per vehicle, the lowest value among all seven evaluated methods, and limited energy above the safe terminal reserve to 108 kWh per vehicle. Its charging events were distributed across five stations, with individual station shares between 15% and 23%. Taken together, these results show that prospective action modelling and shared-resource coordination can deliver fast, economical and operationally coherent eHGV fleet control.

# Acknowledgements

This work was supported in part by the Clean Energy and Equitable Transportation Solutions (CLEETS) NSF-UKRI Global Center under NSF Award No. 2330565 and UKRI Award No. EP/Y026233/1.